\documentclass[twocolumn, tighten, times]{aastex61}
\usepackage{graphicx}
\usepackage{epstopdf}
\usepackage{float}

\usepackage{color}
\usepackage{soul}

\renewcommand\baselinestretch{1.3}

\shorttitle{Unsupervised morphological classification of galaxies}

\shortauthors{Zhou et al.}

\begin{document}

\title{Automatic morphological classification of galaxies: convolutional autoencoder and bagging-based multiclustering model}

\correspondingauthor{Guanwen Fang}
\email{wen@mail.ustc.edu.cn}

\author{ChiChun Zhou}
\altaffiliation{Yizhou Gu and ChiChun Zhou contributed equally to this work}
\affil{School of Engineering, Dali University, Dali 671003, People's Republic of China}

\author{Yizhou Gu}
\affil{School of Physics and Astronomy, Shanghai Jiao Tong University, 800 Dongchuan Road, Minhang, Shanghai 200240, People's Republic of China}

\author{Guanwen Fang}
\affil{Institute of Astronomy and Astrophysics, Anqing Normal University, Anqing 246133, People's Republic of China; wen@mail.ustc.edu.cn}
\author{Zesen Lin}
\affil{CAS Key Laboratory for Research in Galaxies and Cosmology, Department of Astronomy, University of Science and Technology of China, Hefei 230026, People's Republic of China}
\affil{School of Astronomy and Space Science, University of Science and Technology of China, Hefei 230026, People's Republic of China}

\begin{abstract}
In order to obtain morphological information of unlabeled galaxies, we present an unsupervised machine-learning (UML) method for morphological classification of galaxies, which can be summarized as two aspects: (1) the methodology of convolutional autoencoder (CAE) is used to reduce the dimensions and extract features from the imaging data; (2) the bagging-based multiclustering model is proposed to obtain the classifications with high confidence at the cost of rejecting
the disputed sources that are inconsistently voted. We apply this method on the sample of galaxies with $H<24.5$ in CANDELS. Galaxies are clustered into 100 groups, each contains galaxies with analogous characteristics. To explore the robustness of the morphological classifications, we merge 100 groups into five  categories by visual verification, including spheroid, early-type disk, late-type disk, irregular, and unclassifiable. After eliminating the unclassifiable category and the sources with inconsistent voting, the purity of the remaining four subclasses are significantly improved. Massive galaxies ($M_*>10^{10}M_\odot$) are selected to investigate the connection with other physical properties. The classification scheme separates galaxies well in the U-V and V-J color space and Gini--$M_{20}$ space. The gradual tendency of S\'{e}rsic indexes and effective radii is shown from the spheroid subclass to the irregular subclass. It suggests that the combination of CAE and multi-clustering strategy is an effective method to cluster galaxies with similar features and can yield high-quality morphological classifications. Our study demonstrates the feasibility of UML in morphological analysis that would develop and serve the future observations made with China Space Station telescope.
\end{abstract}


\section{Introduction}

Galaxy morphology is an important characteristic relevant to other key physical properties,
such as the stellar mass, the color, the star-formation rate (SFR),
the gas content, the galaxy age, and the environment (e.g., \citealt{1980ApJ...236..351D, 2003MNRAS.341...54K,
2004MNRAS.353..713K, 2014MNRAS.440..843O, 2014MNRAS.440..889S, 2017ApJ...847..134K, 2018ApJ...855...10G, 2019A&A...631A..38L}).
The morphological diversity of galaxies implies the difference in the evolutionary histories of galaxies.
As the development of telescopes and instruments advances, the automatic morphological analysis and
classification of galaxies are imperiously demanded to help understand galaxy formation rate and evolution.
To exploit the potentials of galaxy morphologies from current and future surveys, we
present an unsupervised machine-learning (UML) method for the morphological classification through the astronomical imaging data.

The apparent visual morphology might be the earliest measurement of galaxies \citep{1926ApJ....64..321H}.
Human classifiers can judge the morphological type of galaxies by a certain classification
scheme (e.g, \citealt{1926ApJ....64..321H, 1959HDP....53..275D, 1960ApJ...131..558V}).
Large projects can employ many human classifiers to give the  label or probability of galaxy morphology (e.g.,  \citealt{2011MNRAS.410..166L}). Indeed, conventional visual classification may have some incompatible results due to subjective deviation. However, what's more important is that conventional visual classification is quite low inefficiency since galaxies should be inspected one by one.

Except the visual classifications by eyes, many techniques are developed to quantify the galaxy
structure and the morphology. To obtain the morphological classifications of galaxies, the key is to extract the morphological features from the raw images. Parametric measurements describe the light profiles of galaxies
using the mathematical models with a set of parameters (e.g. \citealt{1963BAAA....6...41S, 2012ApJS..203...24V, 2014ApJ...788...11L}).
Nonparametric measurements design several model-independent parameters to describe some characteristics of galaxy morphologies
(e.g. \citealt{1994ApJ...432...75A, 2003ApJS..147....1C, 2004AJ....128..163L, 2013MNRAS.434..282F};
Also see the review by \citealt{2014ARA&A..52..291C}).
Relying on the parameter spaces constituted of the morphological features, several approaches are
proposed to distinguish the morphological types, such as the concentration, asymmetry, smoothness
method \citep{2014ARA&A..52..291C}, the principal component analysis \citep{2007ApJS..172..406S},
and the support-vector machine \citep{2008A&A...478..971H}. The structural parameters reduce the complexity of morphological descriptions.
However, it rejects the abundant information hidden in all pixels and may lead to failures of morphological distinction in some cases.

The convolutional neural network is one of the most popular methods of supervised machine learning applied
in astronomy to classify galaxy morphologies automatically (e.g.,  \citealt{ 2015ApJS..221....8H, 2015MNRAS.450.1441D, 2019MNRAS.483.2968W}).
It can extract enormous amount of information contained in the pixels themselves hierarchically and
have a good performance in mimicking the human perceptions with a high efficiency. However, it is
a supervised-learning method which means that it is highly dependent on the prelabeled training data set.
By training the model on labeled data,
the supervised-learning method is good at mimicking human perceptions. 

The methodology of UML might be another way to the morphological classification,
which does not need a prelabeled training set labeled by human classifiers. Hence it has no subjective
deviations of humans and may exploit a new angle of galaxy morphologies from the machine's view.
The methodology of UML is ideally suited to the morphological analysis of Big Data surveys, which
has been applied to the automatic classification of optical/NIR and radio images (e.g., \citealt{2019PASP..131j8011R, 2020MNRAS.497.2730G}).
\cite{2018MNRAS.473.1108H} and \cite{2020MNRAS.494.3750C} apply the growing neural gas
algorithm \citep{Fritzke95} to extract features from images and the hierarchical clustering
technique is responsible for gathering the galaxies with similar features.
The convolutional autoencoder (CAE) \citep{masci2011stacked} is also another effective technique
to extract features from images. The combination of autoencoder and clustering algorithm is able
to obtain the reasonable results on strong lensing identification \citep{2020MNRAS.494.3750C} and
morphological classification \citep{2021MNRAS.tmp..718C}.

The upcoming China Space Station Telescope (CSST) will image the sky in 7 bands ($NUV, u, g, r, i, z$, and $y$) and vastly enrich the photometric data with a wide survey covered 17,500 deg$^2$ with 5$\sigma$ depth of $r = 26.0$ mag and a deep survey covered 400 deg$^2$ with 5$\sigma$ depth of $r = 27.2$.
To prepare for and to exploit the imaging data of the CSST, we plan to develop an UML method for galaxy classifications.

In this paper, we pioneer in the use of unsupervised classification in the five CANDELS fields. The CAE is trained for the extraction of unsupervised features from the raw imaging data.
 After that, to avoid the bias from one single clustering algorithm, a bagging-based multiclustering method is proposed which considers the results of three different clustering algorithms.
Although at the cost of eliminating the disputed sources, galaxies are clustered into 100 groups under a comprehensive definition of similarity. As a result, the purity in each classification is significantly improved.
To test the feasibility of our method, we merge 100 groups into five subclasses by visual verification, including spheroid (SPH), early-type disk (ETD), late-type disk (LTD), irregular (IRR), and unclassifiable (UNC).  After discarding the disputed sources and the UNC category, we investigate the connection with colors and morphological parameters using the massive galaxies ($M_*>10^{10} M_\odot$).
Moreover, by using the t-SNE visualize technique \citep{van2008visualizing}, the comparisons between our result and that of CANDELS visual
classifications \citep{2015ApJS..221...11K} and
the supervised deep-learning method \citep{2015ApJS..221....8H} are given. It suggests that the proposed method, combination of CAE and multiclustering strategy, is an effective method to cluster galaxies with similar features and can yield high-quality morphological classifications, which are useful in other downstream tasks.

This paper is organized as follows. The data set and our sample construction are described in Section \ref{sec:data}.
The method is described in Section \ref{sec:method}, which includes the CAE and the bagging-based multiclustering method.
We compare our clustering result with other physical properties of massive galaxies and the results of other works in Section \ref{sec:result}.
Main conclusions and outlooks are summarized in Section \ref{sec:sum}.
When converting the effective radii of galaxy from observational scales (arcsec) to physical scales (kpc), we assume the cosmological parameters as following: $H_0=70\,{\rm km~s}^{-1}\,{\rm Mpc}^{-1}$, $\Omega_m=0.30$, $\Omega_{\Lambda}=0.70$.

\section{Data and sample selection} \label{sec:data}
CANDELS have provided WFC3 and ACS photometry over $\sim$ 900 arcmin$^2$ in five fields: AEGIS, COSMOS, GOODS-N, GOODS-S, and UDS \citep{2011ApJS..197...35G, 2011ApJS..197...36K}. The 3D-HST Treasury Program provides a large amount of the data products, refering to photometry \citep{2014ApJS..214...24S} and grism spectra \citep{2016ApJS..225...27M}. \cite{2016ApJS..225...27M} provide a ``best'' redshift catalog by merging their grism-based results with the photometric results in \cite{2014ApJS..214...24S}.

Here, we give a brief introduction of the 3D-HST data set.
The redshift of a galaxy is arranged in the order of spectroscopic redshift, grism redshift, and photometric redshift.
If spectroscopic redshift is not available, then we adopt grism redshift. If grism redshift is not available, we use photometric redshift instead.
Photometric redshifts are derived by the spectral Energy Distributions (SEDs) ranging from 0.3 to 8.0 $\mu$m with the EAZY code (\citealt{2008ApJ...686.1503B}). It applies the linear combination of seven galaxy spectral templates, wihch are the defaults in \cite{2008ApJ...686.1503B}, including the five templates from the library of P\'{E}GASE stellar population synthesis models (\citealt{1997A&A...326..950F}), one young, dusty template and one old, red galaxy template (see also \citealt{2011ApJ...735...86W}). The error of photometric redshift  reaches $\Delta z /(1+z) \approx 0.02$ on average. Grism redshifts are determined by the combined  fitting of spectrum and photometry data, using a modified EAZY code. For most galaxies, their grism redshifts are of extremely high accuracy with $\Delta z /(1+z) \approx 0.003$. Furthermore, spectroscopic redshifts are compiled from previous literature in the five fields (see \cite{2014ApJS..214...24S} for detail field by field).  
The rest-frame colors are derived with the EAZY code at the same time. In addition, stellar masses are estimated with the FAST code (\citealt{2009ApJ...700..221K}), by assuming exponentially declining star-formation histories, solar metallicity, \cite{2000ApJ...533..682C} dust attenuation law and \cite{2003MNRAS.344.1000B} stellar population synthesis models with a \cite{2003PASP..115..763C} initial mass function.

\begin{figure}
\centering
\includegraphics[scale=0.8]{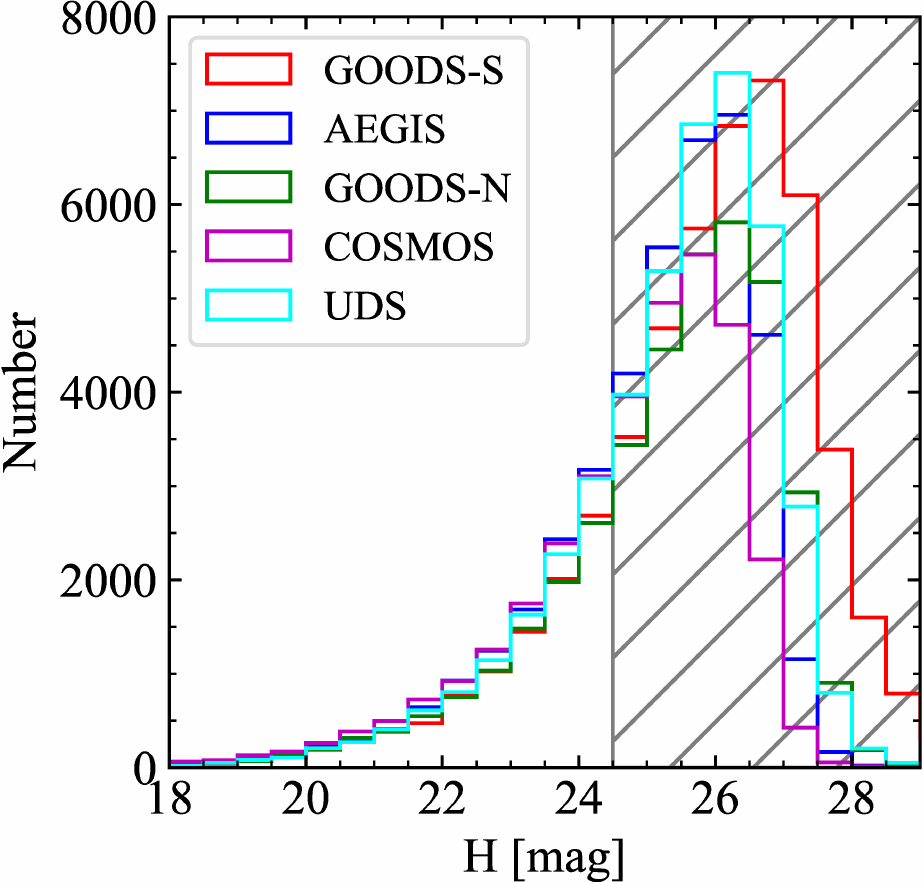}
\caption{H-band magnitude distribution in each of the five fields. A magnitude limit of F160W $<$ 24.5 is imposed. }
\label{fig01}
\end{figure}

The detailed morphological classification depends on the spatial resolution. In this paper, we use the HST/F160W (H-band) selected catalogs (v4.1.5) and H-band images  from the 3D--{\it HST} project  \footnote{\url{http://3dhst.research.yale.edu/}} \citep{2014ApJS..214...24S, 2016ApJS..225...27M}. We select all those galaxies with F160W $<$ 24.5 mag,
ensuring the galaxies being bright enough to obtain the reliable morphologies. An additional criterion
is the flag {\tt use\_phot=1}, which means that the object (1) is not too faint and not a star,
(2) is not contaminated by a bright source, (3) is well exposed both in the
F125W and F160W, (4) has a signal-to-noise ratio (S/N) $>$ 3 in F160W images, and
(5) has “noncatastrophic” photometric redshift and stellar population fits \citep{2014ApJS..214...24S}. After removing the images containing abnormal values due to the bad pixels or locating in survey boundaries, our initial sample contains 47149 galaxies with a median redshift of $\sim$ 1.1.
Figure \ref{fig01} shows the distributions of H-band magnitudes in each of five fields.

\section{The unsupervised method for morphological classification}
\label{sec:method}

\begin{figure*}
\centering
\includegraphics[scale=0.45]{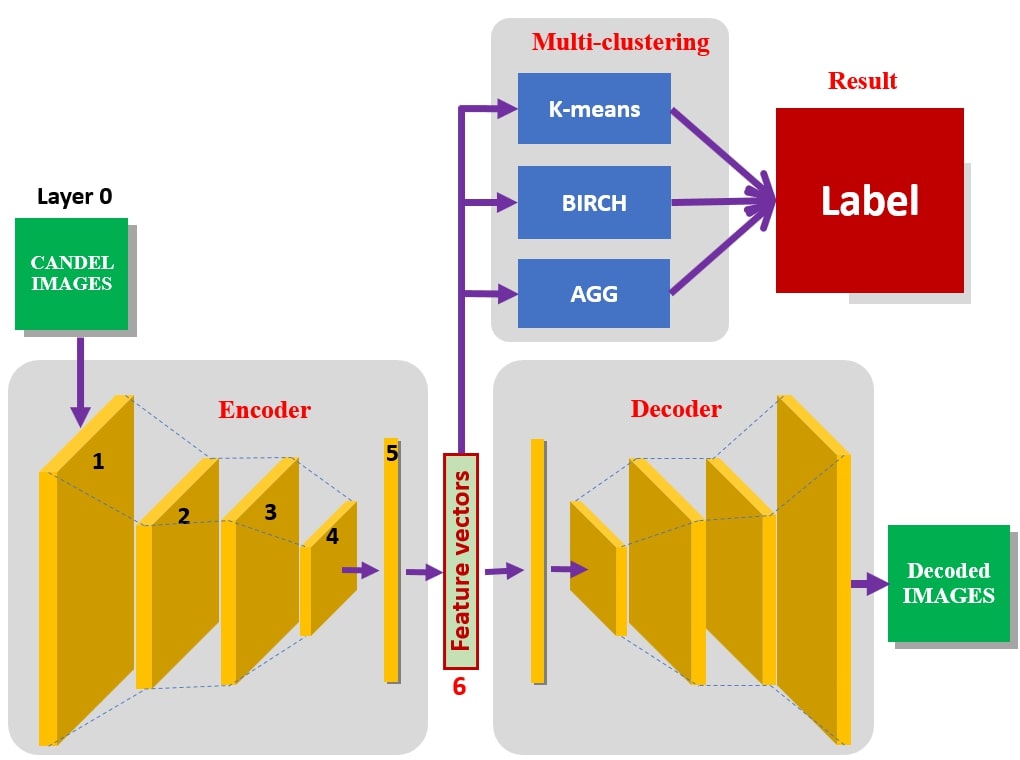}
\caption{Schematic diagram of the convolutional autoencoder architecture and multi-clustering strategy. The CAE is firstly trained by minimizing the square mean error of pixels of raw images and decoded images (Eq. \ref{loss}). Then, based on the encoded feature vectors given by the CAE, three clustering models vote on the labels of galaxies.
}
\label{fig02}
\end{figure*}

In this section, we introduce the main process for morphological classification.
Our unsupervised method consists of two steps: feature extraction and multiclustering.
Figure \ref{fig02} gives an overview of the CAE architecture and multiclustering strategy.
Firstly, we use the CAE to compress the dimension and extract morphological information of galaxies from the raw image. Then, we cluster galaxies into 100 groups by bagging-based multiclustering methods: three clustering methods vote on  accepting or rejecting of a galaxy in a given group.
The bagging-based multiclustering methods assures the classifications with a high degree of confidence, but at the cost of rejecting the disputed sources that are inconsistently voted.

\subsection{Data preprocessing}

Raw imaging data with large dimensions usually contain noises and useless information, although it contains all morphological information of galaxies. In this paper, the size of raw imaging data is $28\times28$ with a source in the center, corresponding to $1^{\prime \prime}.68 \times 1^{\prime \prime}.68$. Processing raw pixels of imaging data directly may have negative effects on the downstream tasks and is computing resources consuming. In order to avoid such effects and reduce the calculation amount, a max--min normalization pretreatment is applied to each cutout.
Namely, the flux of each pixel in float type is converted to
nonnegative integers by the following procedure:

\begin{equation}\label{eq:normalization}
{\rm new~pixel}=\left \lfloor \frac{{\rm old~pixel}-m}{M-m} \times N\right\rfloor,
\end{equation}
where $M$ and $m$ are the maximum and minimum fluxes of pixels, respectively,
and $\left\lfloor x\right\rfloor $ gives the largest integer that is smaller than $x$,
for example $\left\lfloor 3.5\right\rfloor =3$.
$N$ takes 500 in the present paper, which is a given integer describing the degree of discretization.
The choice of $N$ would not significantly change our results. We show in Section \ref{subsec:diff_paras} that $N=500$ is a reasonable choice since the final classification is nearly unchanged when another value of $N$ is adopted.

\subsection{Dimensionality reduction by CAE}
\label{subsec:dim_cae}

The CAE is a convolutional network involving the operations of convolution and pooling, which can extract informations from images effectively \citep{masci2011stacked, krizhevsky2012imagenet}.
It is dedicated to extract information from images and thus compress the dimension of features of photometric data hierarchically \citep{ng2011sparse,wang2016auto}.

In the CAE, operations of convolution and pooling encode the pixels and give an encoded feature with lower dimension. Operations of deconvolution and unpooling decode the encoded feature and reconstruct the pixels.
Table \ref{tab_CAE} summarizes the outline of CAE architecture. In this work, the dimension of the input data is $28\times28$ ($784$).
By using the CAE, we compress the dimension of features from $784$ to a lower number
and improve the efficiency of calculation.  In this work, we choose to use $40$ as the target dimension of the compressed features to balance the effectiveness and efficiency (i.e., the dimension of the hidden layer $dim_{\rm L6}=40$). After fixing the final feature number, the basic concept of the training is to reduce the differences between the input images and the reconstructed images, which is described by the loss function
\begin{equation}
loss=\frac{1}{784n}\sum_{i=1}^{n}\sum_{j,k}^{28}\left(  \hat{y}_{j,k}-y_{j,k}\right)  ^{2},
\label{loss}
\end{equation}
where $n$ is the number of subsamples in a batch, and $y_{j,k}$ and $\hat{y}_{j,k}$ are input and reconstructed pixels in position $j$ and $k$. By tuning the parameters in the networks to minimize the loss function, the CAE learns to encode the raw photometric data, which gives and decodes the compressed features.

\begin{figure*}
\centering
\includegraphics[width=\textwidth]{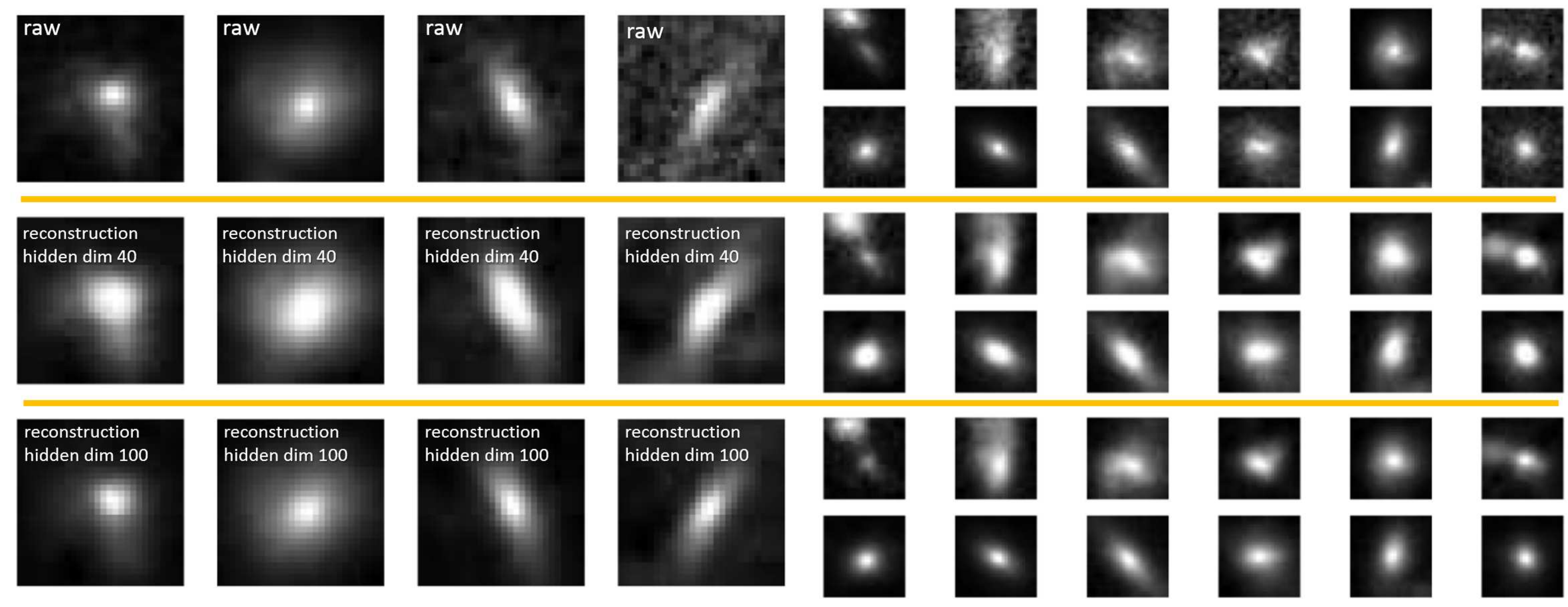}
\caption{Top panels are the demonstrations of raw images of random selected galaxies.
Middle and bottom panels are the reconstructed images using hidden-layer size 40 and 100, respectively. Three panels are separated by the horizontal lines. 
}
\label{fig03}
\end{figure*}

Figure \ref{fig03} illustrates several reconstructed images by CAE with hidden-layer sizes of $dim_{\rm L6}=40$ and 100, together with the corresponding input images for comparison. Clearly, $dim_{\rm L6}=40$ is enough to recover the major morphological information of our galaxies. This method is useful in processing the raw imaging data of galaxies and can be used in other deep survey investigation.

\begin{deluxetable*}{cccccc}
\setlength{\tabcolsep}{5mm}
\tablecaption{Outline of CAE Architecture and Layer Configuration. \label{tab_CAE}}
\tablehead{\colhead{Network Section} &
\colhead{Layer} &
\colhead{Operation} &
\colhead{Dimension} &
\colhead{Filter Size} &
\colhead{Stride}
}
\startdata
Encoder   & L0 & Input       & $28\times28\times1$            & ... & ...\\
          & L1 & Convolution & $28\times28\times128$ & $3\times3$ & ... \\
          & L2 & Maxpooling  & $14\times14\times128$ & $2\times2$ & $2\times2$\\
          & L3 & Convolution & $14\times14\times128$ & $3\times3$ & ... \\
          & L4 & Maxpooling  & $7 \times7 \times128$ & $2\times2$ & $2\times2$\\
          & L5 & Unfolding    & $6272$ & ... & ...\\
          & L6 & Full connection  & $40$ & ... & ...\\
\enddata
\end{deluxetable*}

\subsection{The bagging-based multiclustering models}
\label{subsec:voting}

The clustering method is a class of UML methods that is efficient in handling large amount of unlabeled data and is good at clustering subsamples with similar properties.
In this section, we describe the proposed bagging-based multiclustering strategy, which is considered to be an efficient method obtaining the high-quality subsamples with similar properties automatically.

\begin{figure*}
\centering
\includegraphics[scale=0.5]{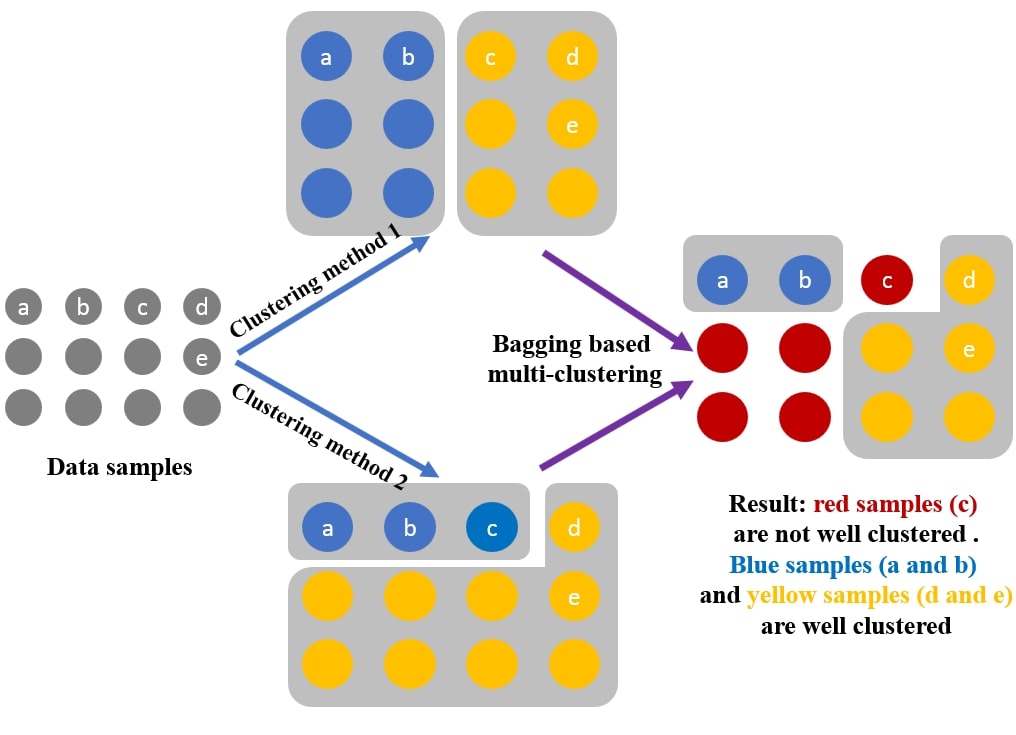}
\caption{An illustration of the  bagging-based multiclustering
models: the method based on voting.  Red samples, such as the sample c,
are not well clustered as the object c is disputed by methods 1 and 2 and thus are eliminated. Blue samples,
such as object a and object b, and yellow samples, such as object d and object e, are well clustered as
they are consistently voted by all the methods.}
\label{fig04}
\end{figure*}

There are different clustering algorithms.
However, we are cognizant of the problem that different clustering algorithms use different similarity
definitions and techniques, possibly resulting in inconsistent clustering output even for the same data set.
For one clustering method, to  analyze similarities between subsamples from a single perspective may lead to misclustering.
Therefore, we propose the bagging-based multiclustering method as shown in Figure \ref{fig04}: subsamples
are finally clustered into one group if they are clustered into one group by all the clustering algorithms.
That is, the strategy aims at providing morphological classifications with high confidence at the
cost of rejecting a number of disputed subsamples.

The key procedure to implement the bagging-based multiclustering strategy is
to align labels generated by different clustering algorithms before voting,
since labels assigned by different algorithms might be different and are meaningless
in direct comparisons. To align labels, we set the labels given by the k-means clustering
algorithm as the primary ones and assign them to every group clustered by the other two
algorithms according to the k-means label with the highest frequency in that group.
After aligning all labels given by all clustered algorithms, voting starts.

Clustering algorithms can be divided into several categories \citep{elavarasi2011survey}:
(1) partition clustering algorithms that
split subsamples into k partitions \citep{elavarasi2011survey},
(2) hierarchical clustering algorithms that
group subsamples into form a tree shaped structure \citep{murtagh1983survey},
(3) density-based clustering algorithms that grow the groups until the density in the neighborhood exceeds certain threshold \citep{han2011data},
(4) spectral clustering algorithms where
groups are formed by partition subsamples using the similarity matrix \citep{meila2001comparison},
and
(5) grid based clustering algorithms that quantize the subsample space into cells with a grid structure \citep{han2011data}.

In this work, our bagging based multi-clustering strategy consists
of three typical clustering algorithms, i.e., the k-means clustering algorithm \citep{hartigan1979algorithm, kanungo2002efficient},
the agglomerative clustering algorithm (AGG; \citealt{murtagh1983survey,murtagh2014ward}),
and the balance iterative reducing and clustering using hierarchies (BIRCH; \citealt{zhang1996birch,zhang1997birch,peng2018balanced}).
In other words, each subsample is firstly voted by the k-means, the AGG,
and the BIRCH algorithms, respectively, then the well clustered subsamples are accepted and grouped, while the disputed subsamples are rejected. These clustering algorithms are described briefly below:
\begin{enumerate}
\item The k-means clustering algorithm is
a typical kind of partition clustering algorithm, which partitions $N$ subsamples into $k$ clusters based on the nearest mean distance.
In this algorithm, subsamples are iteratively processed.
Firstly, the algorithm selects $k$ subsamples randomly as predefined clusters.
Secondly, a new subsample is grouped into one of these predefined clusters by minimizing the distance between the subsample and the center of the cluster.
Thirdly, the center of the cluster is reevaluated after the new subsample is added in.
The process is repeated until no best clusters are found.
\item The AGG algorithm, a typical
kind of hierarchical clustering algorithms, groups subsamples by constructing a tree-based representation of the subsamples. In the AGG algorithm, clustering happens in a bottom-up manner. Firstly,
each subsample is considered as a singleton cluster or so called the leaf. Secondly, two similar clusters
are combined into a new bigger cluster or so called the node. This procedure is iterated until all subsamples are member of one single cluster or so called the root.
\item The BIRCH algorithm,
another typical kind of hierarchical clustering algorithms, summarizes a cluster of subsamples by using notions of clustering feature and represents a cluster hierarchy by using clustering feature tree.
In the BIRCH algorithm, the clustering feature
is essentially a summary of the statistics for the given cluster \citep{han2011data}
and a clustering feature tree is a height-balanced tree that stores the clustering features for a hierarchical clustering \citep{han2011data}.
By scanning the entire database, BIRCH builds an initial clustering feature tree.
Then, in clustering the leaf nodes of the clustering feature tree, sparse clusters are removed as outliers and dense clusters are grouped into larger one.
\end{enumerate}

The bagging of three typical kinds of clustering algorithms is executed in the multiclustering block of Fig \ref{fig02}.
Although taking the intersection would reduce the sample size,
it enables one to use a comprehensive similarity definition and provides a new classification scheme of galaxies with high confidence.
This high-quality clustering subsamples can be used as a
training sample in other downstream tasks such as the supervised machine-learning tasks.

\section{RESULTS AND DISCUSSION}
\label{sec:result}

\subsection{Results of morphology classification}
\label{subsec:res_morph}

\begin{figure*}
\centering
\includegraphics[width=\textwidth]{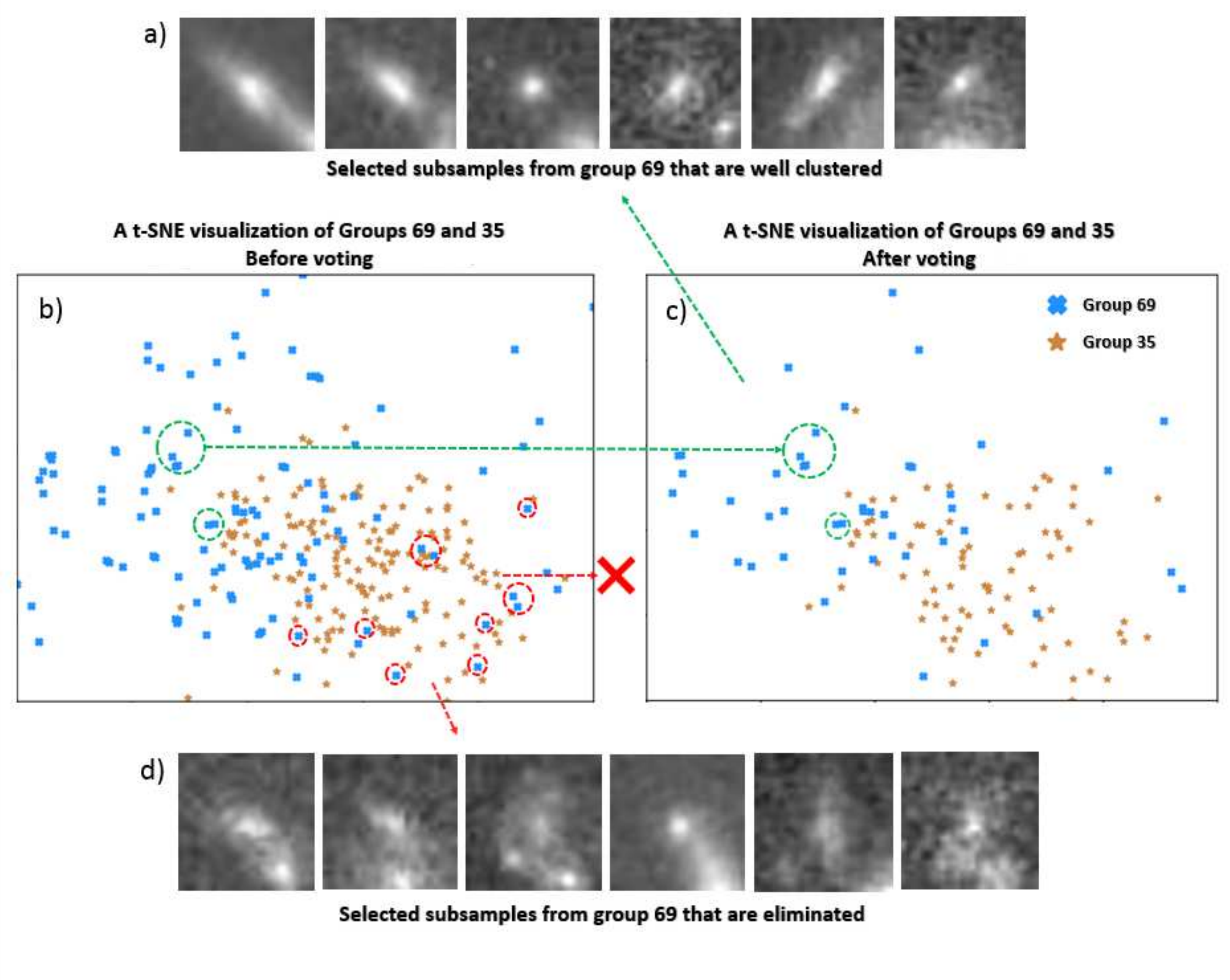}
\caption{ T-SNE diagrams for Groups 35 and 69 before (Panel b) and after (Panel c) the voting. Examples that are well-clustered and are eliminated due to inconsistent voting results by different clustering algorithms are marked out by green and red dashed circles, respectively. The H band cutouts of these examples are also given.}
\label{fig05}
\end{figure*}

Based on the classification scheme described in Section \ref{sec:method}, we classify galaxies with consistent voting by the three clustering algorithms into 100 groups. This strategy improves the purity of each group and provides a high-quality clustering subsamples at the cost of eliminating 22,249 ($\sim 47\%$) disputed galaxies that are inconsistently voted in the voting model.
In this approach, we mainly focus on the remaining 24,900 ($\sim 53\%$) galaxies that are consistently voted.
The excluded subsample will be considered in further studies. For example, by considering the well-clustered galaxies as pre-labeled training data,
we intend to develop supervised methods to group the remaining galaxies.
For the sake of clarity, we eliminate these disputed galaxies from our following discussion.
In this work, we use the t-Distributed Stochastic Neighbor Embedding (t-SNE), which is a technique that visualizes high-dimensional data by giving each datapoint a location in a two- or three-dimensional map \citep{van2008visualizing}, to illustrate the clustering results. In Figure \ref{fig05}, we show the t-SNE diagrams before and after the voting for two groups (i.e., groups 35 and 69). Examples of sources that are well-clustered and are eliminated are marked out, and their cutouts in H band are also presented.

To further investigate the connections between galaxy morphology and other galaxy properties,
we merge the 100 groups into five subclasses, which are 6335 spheroids (SPHs), 3916 early-type disks (ETDs),
4333 late-type disks (LTDs), 9851 irregulars (IRRs), and 465 unclassifiable (UNC) sources, by visual inspection of images  from each group.
SPH galaxies are bulge-dominated, whereas LTD galaxies are disk-dominated.
ETD galaxies are predominantly bulge-dominated, disk galaxies.
IRR subclass includes galaxies with irregular structures or merger evidences.
The remaining groups that cannot be attributed to any one of the above types are collected as UNC sources. The UNC subclass accounts for only $<1\%$ of our sample.

\begin{figure*}
\centering
\includegraphics[width=\textwidth]{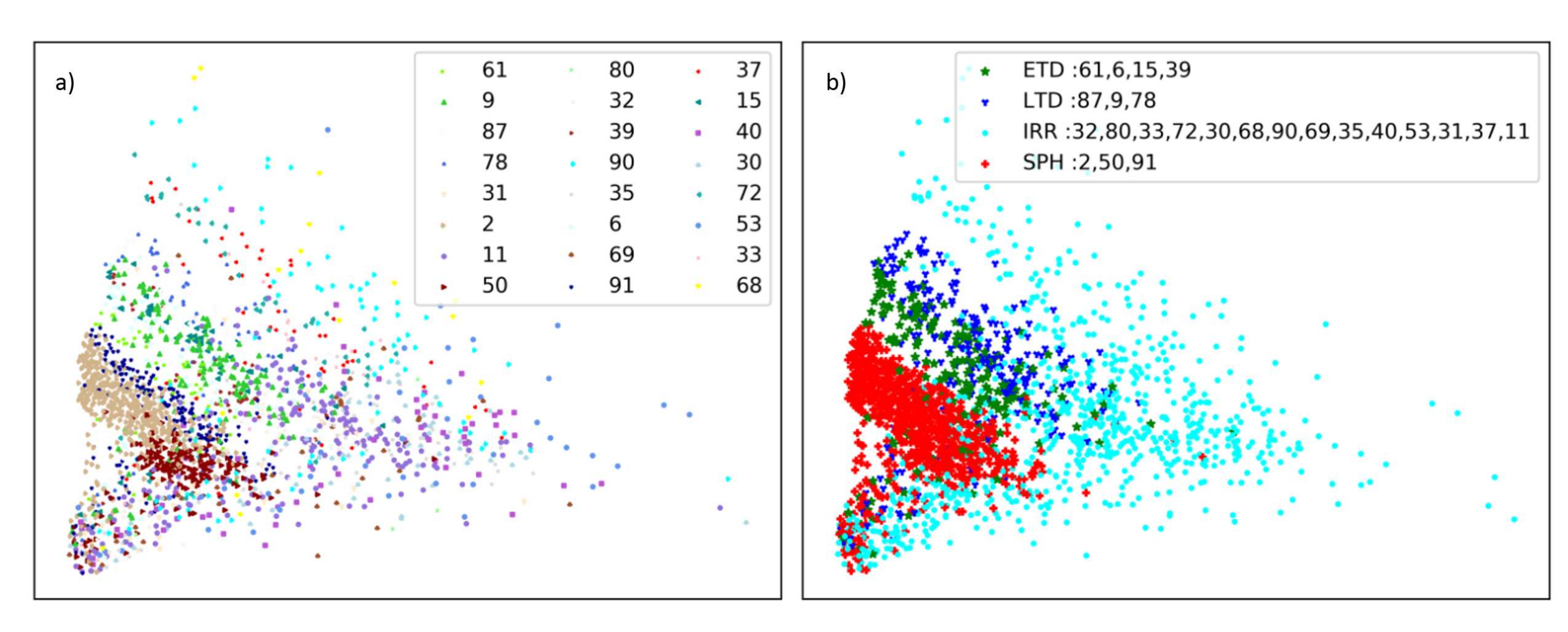}
\caption{Illustration of the procedure of visual merger, showing the t-SNE visualization graph based on raw images of 24 randomly selected groups. Panel (a) is color-coded by the group IDs. Panel (b) is highlighted in different colors, representing the four merged groups by visualization (red: SPH; green: ETD; blue: LTD; cyan: IRR).
}
\label{fig06}
\end{figure*}

\begin{figure*}
\centering
\includegraphics[scale=0.55]{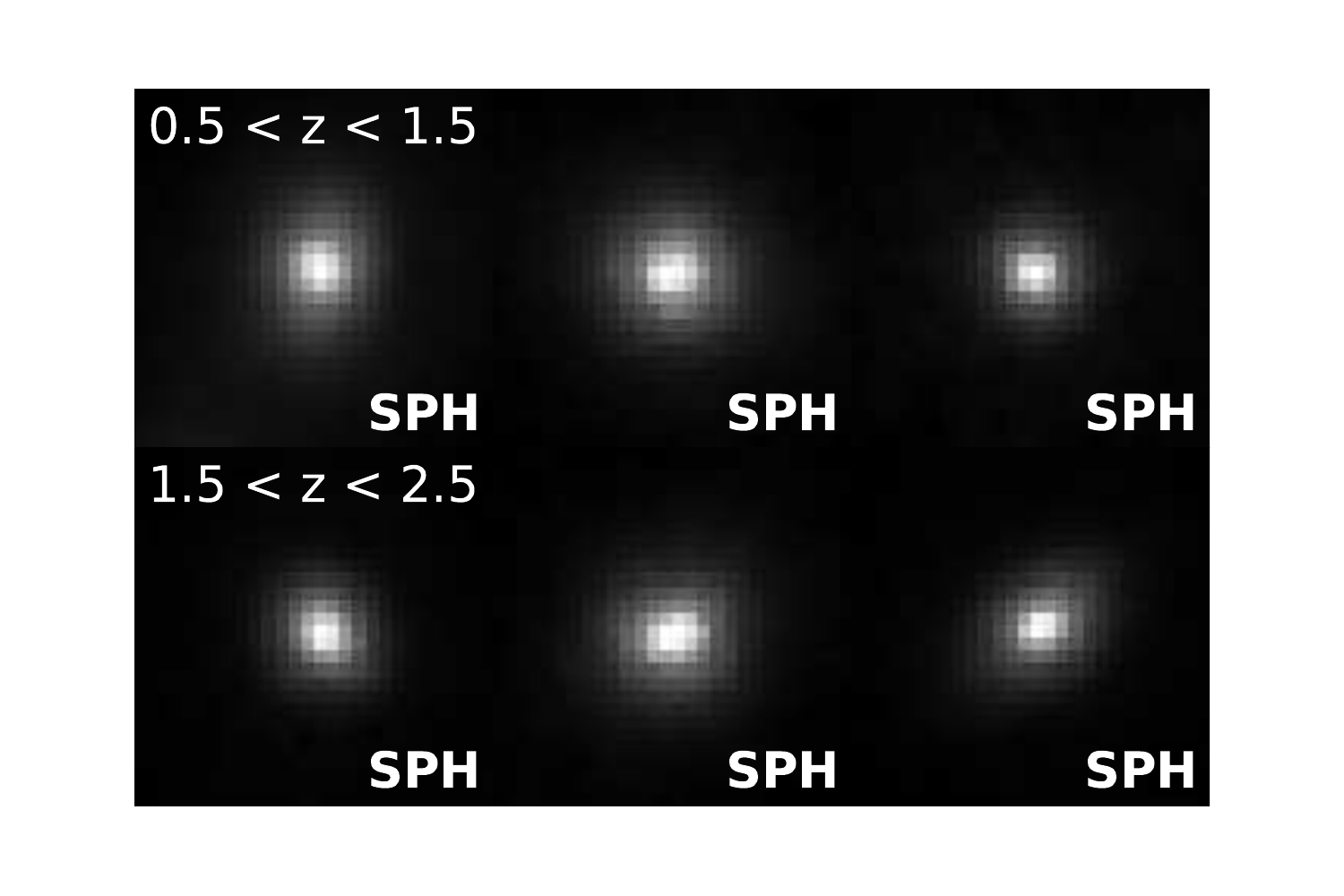}
\includegraphics[scale=0.55]{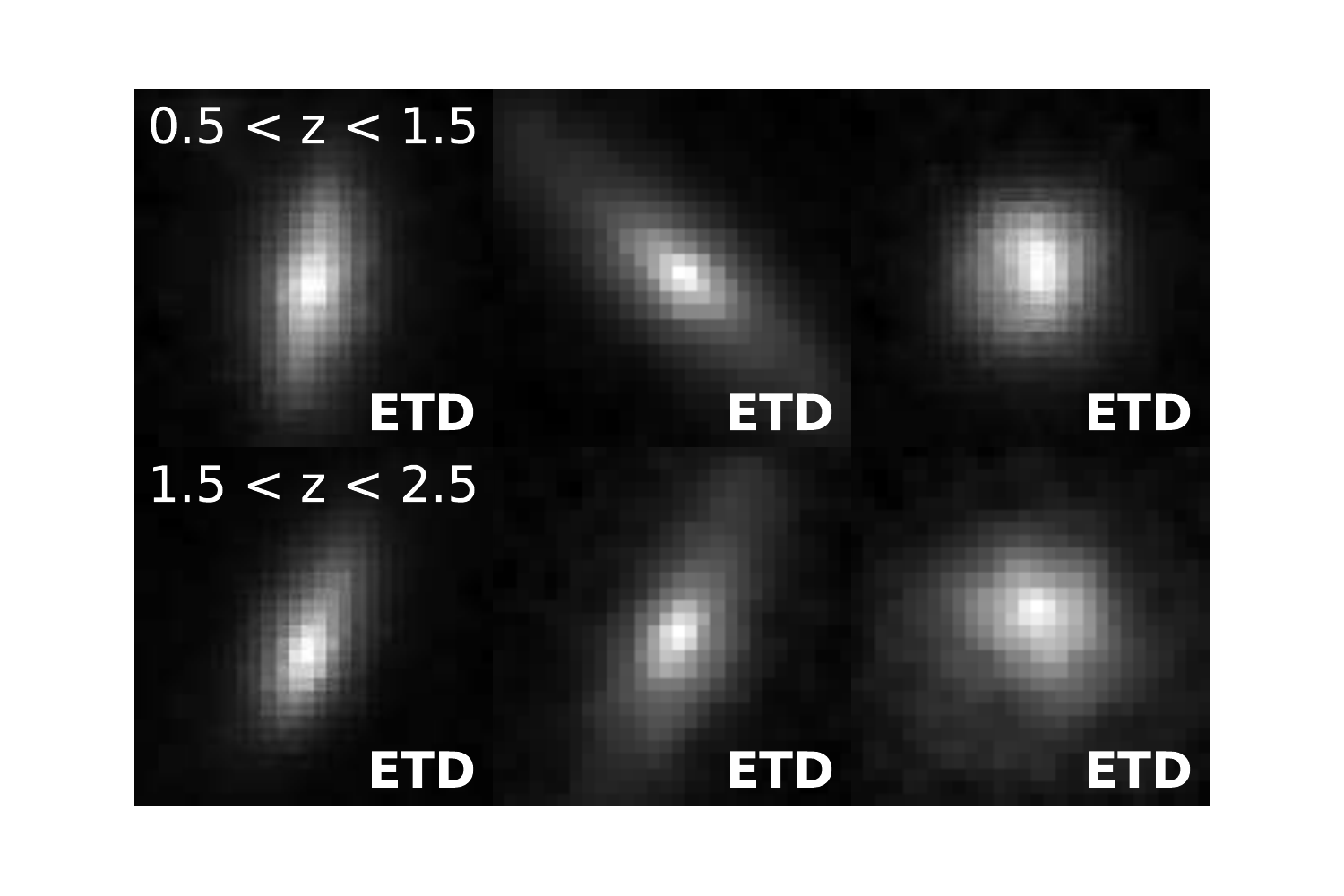}
\includegraphics[scale=0.55]{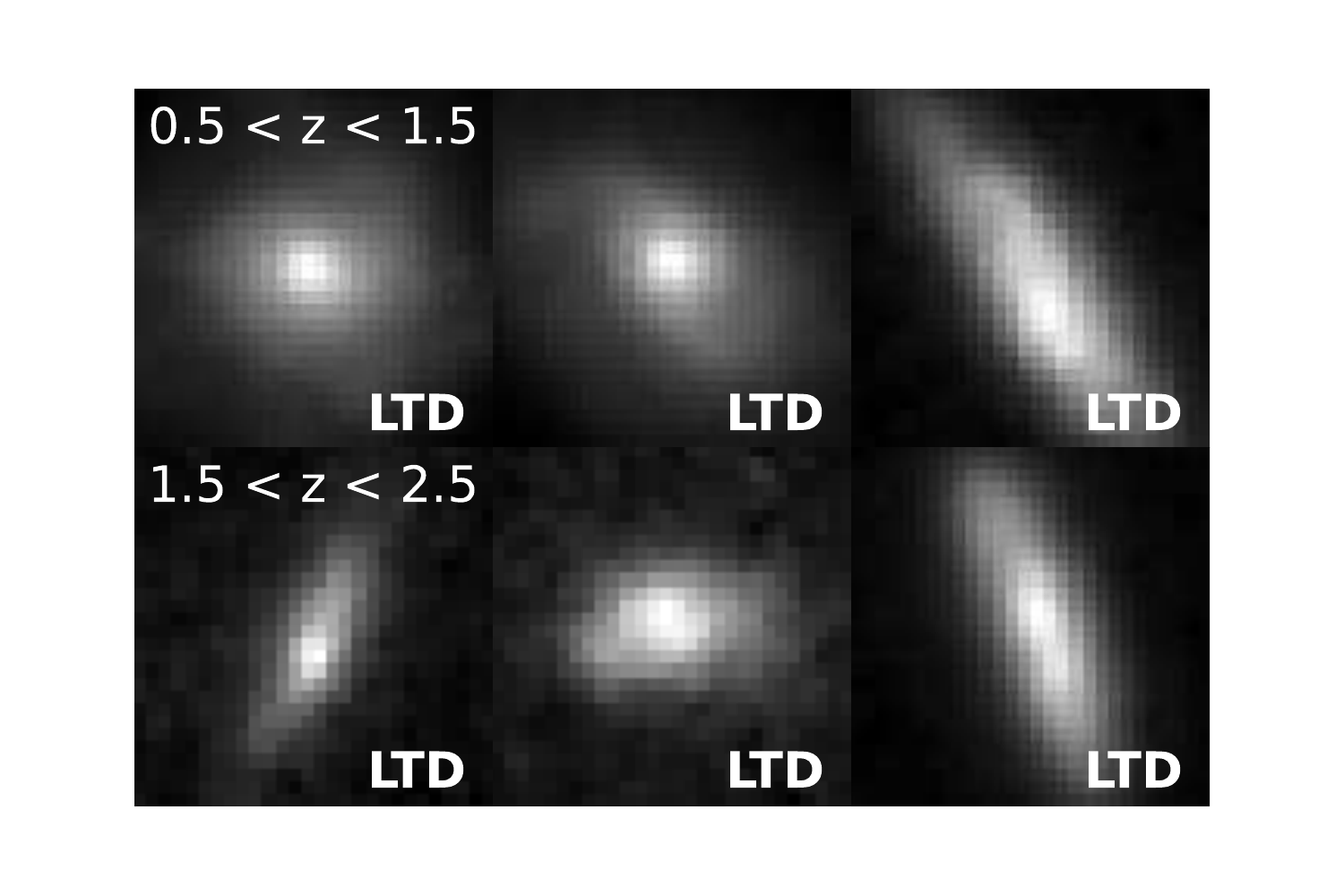}
\includegraphics[scale=0.55]{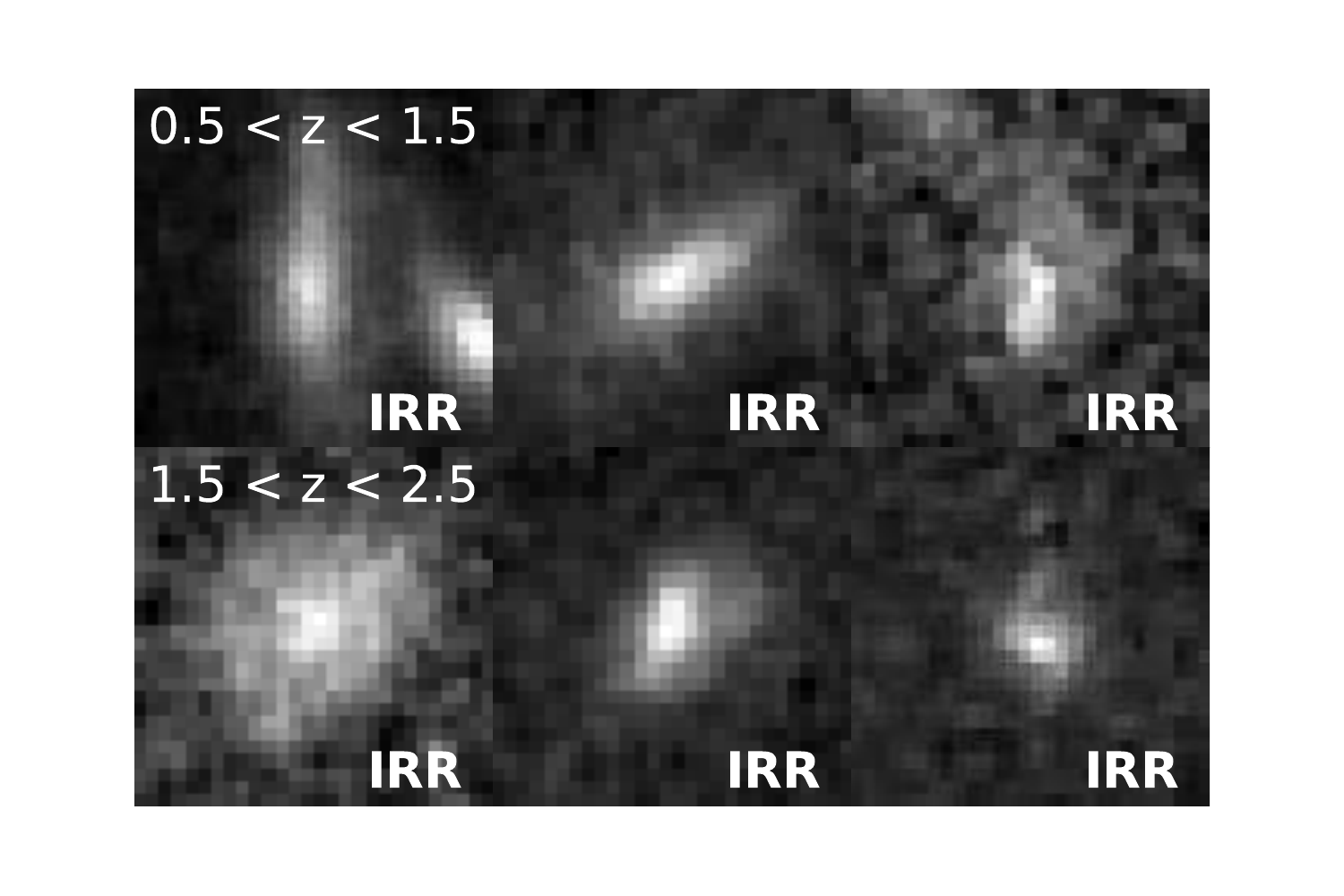}
\includegraphics[scale=0.55]{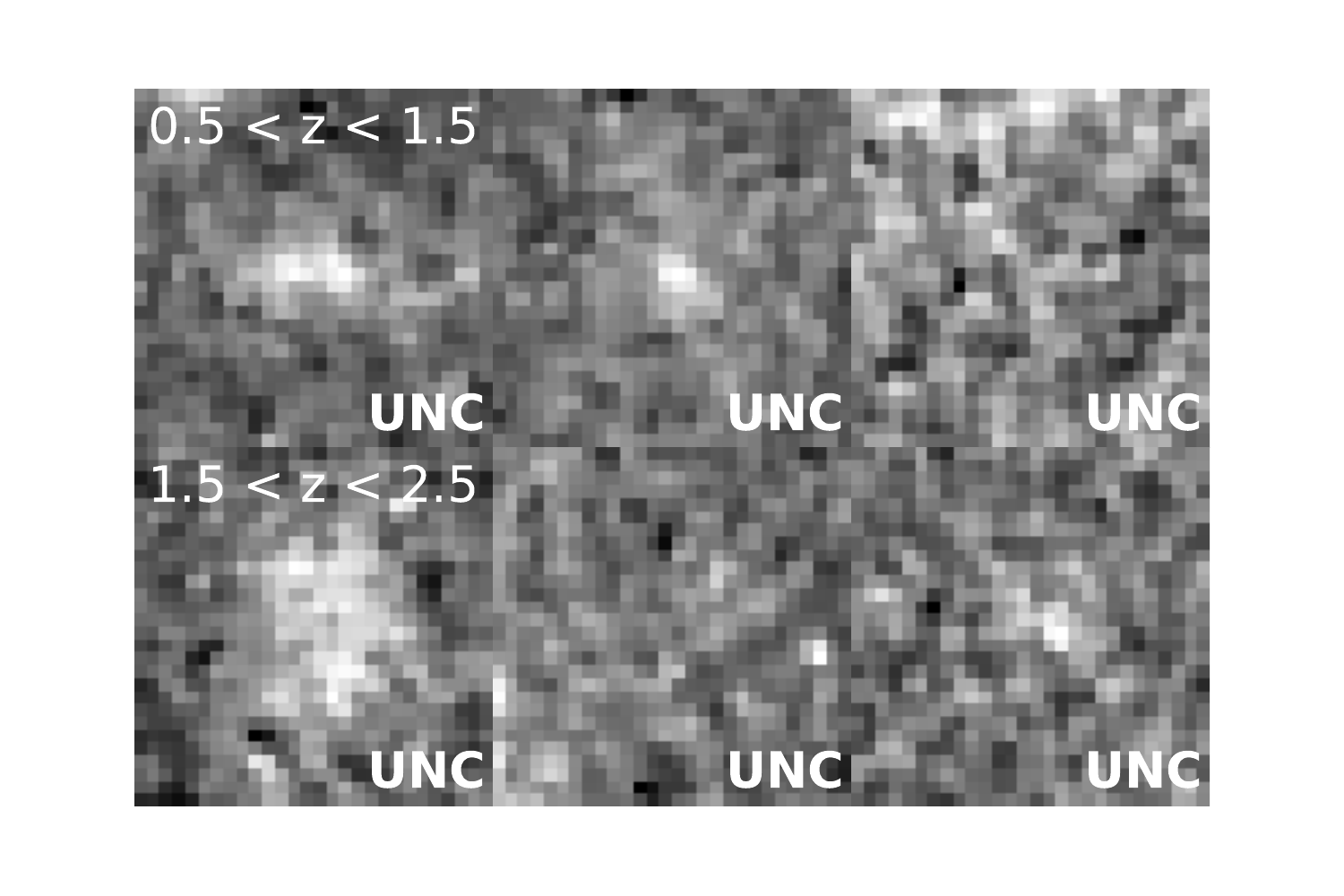}
\caption{
The stamps of galaxies in five subclasses (SPH, ETD, LTD, IRR, and UNC). Six random images are shown in two redshift bins, which illustrate that our method is capable of linking the clustered galaxies to the morphology types. The UNC category is ignored in the following discussion due to its low signal-to-noise ratio.
}
\label{fig07}
\end{figure*}

In order to illustrate the procedure of visual merger, we select 24 groups randomly and exhibit their distributions in the t-SNE graph based on raw images before and after the merger in Figure \ref{fig06}. Due to the small number of UNC sources, none of the groups belonging to this subclass are selected. Obviously, groups merged into the same subclass also tend to be clustered before the merger step expect for the IRR subclass, supporting the reliability of our visual inspection. The F160W cutouts of random examples at $0.5<z<2.5$ for the five subclasses are shown in Figure \ref{fig07}. It is noteworthy that UNC sources seem to have relatively low S/N compared to the other four subclasses. This tendency is confirmed by further examination of the S/N distributions, which reveals that the majority of the UNC population has S/N $\lesssim10$ in H band, while the S/N in H band for most of the galaxies in other subclasses are $>10$. Therefore, we conclude that the unclassifiable feature of UNC sources is mainly due to their low S/N. We will ignore this small population in the following analyses.

Given these merged subclasses, we are able to study the distributions of key galaxy properties as a function of morphological type, as a test of the reliability of our classifications. Since there exists general correlations between morphologies and other physical properties of massive galaxies ($M_*>10^{10}M_\odot$; e.g., \citealt{2008MNRAS.383..907B, 2018ApJ...855...10G}),  we utilize massive galaxies to explore the robustness of the morphological classifications.

\subsection{The distributions in the color--color diagram}%

Observations reveal the morphology--color relation, especially at a fixed stellar mass, (e.g., \citealt{2006MNRAS.368..414D, 2009ApJ...699..105C, 2011ApJ...736..110M, 2012ApJ...751L..44W}) and the morphology--SFR correlation (e.g., \citealt{2012ApJ...751L..44W, 2014MNRAS.441..599B, 2014MNRAS.440..843O, 2015MNRAS.448..237W, 2017MNRAS.465..619B}).
The relationship between the galaxy morphology and the color (or, alternatively, star formation) might be caused by the different properties of bulges and disks \citep{2006MNRAS.368..414D}, while the bulge is considered as a reliable predictor of the state of quiescence \citep{2014ApJ...788...11L, 2016MNRAS.457.2086T}.

\begin{figure}
\centering
\includegraphics[scale=0.6]{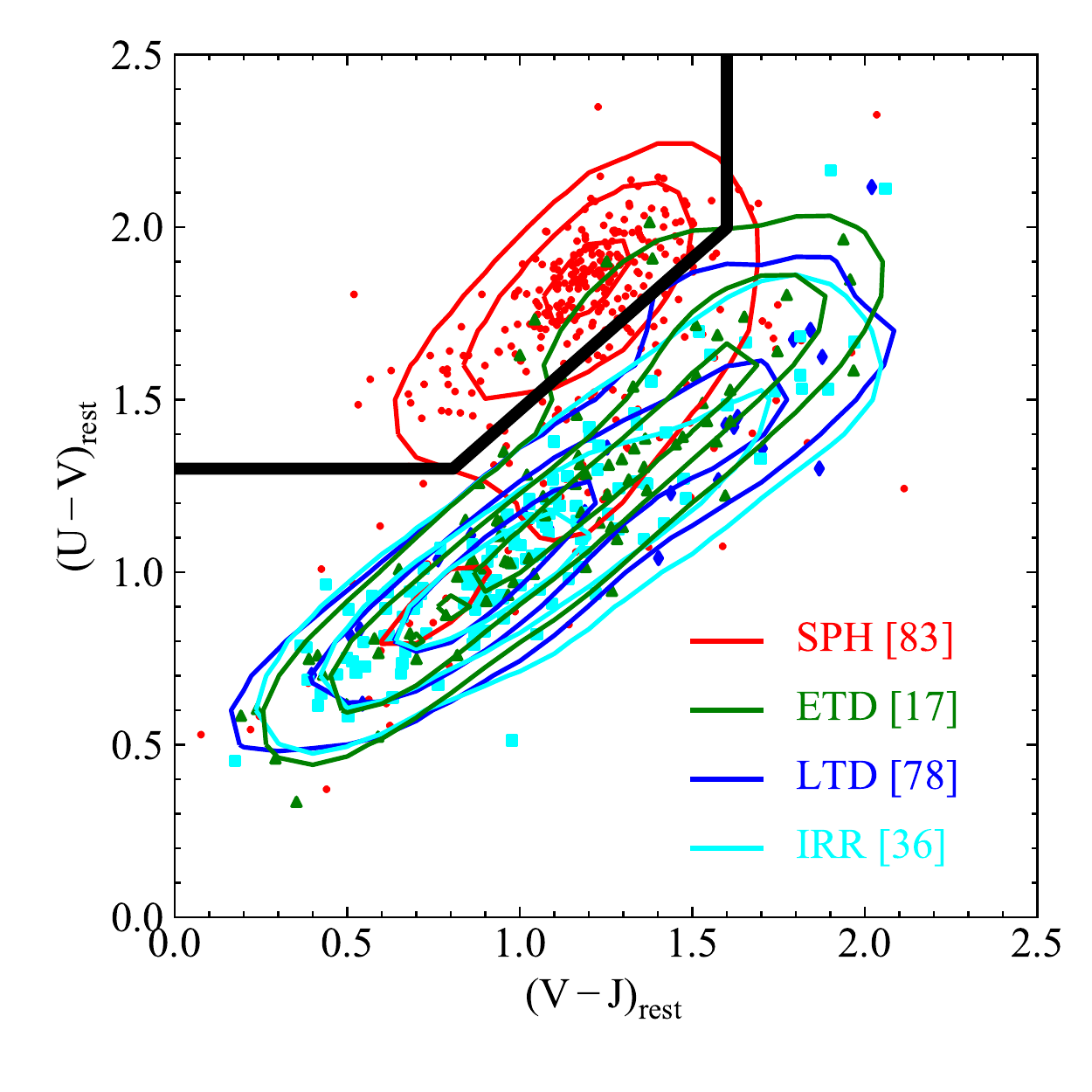}
\caption{Rest-frame $U-V$ versus $V-J$ colors for galaxies that are merged into SPH (red), ETD (green), LTD (blue), and IRR (cyan).  The contour levels correspond to the 20\%, 50\%, and 80\% of the specified galaxies from inner to outskirts.
The distribution of some individual groups is overlapped, where the group id is marked in the square brackets. The SPH, ETD, LTD, and IRR categories have the marker differing, using red points, blue diamonds, green triangles, and cyan squares, respectively. 
}
\label{fig08}
\end{figure}

The {\it UVJ} diagram is a useful diagnosis to separate galaxies into star-forming and quiescent galaxies up to $z = 2.5$ (e.g., \citealt{2009ApJ...691.1879W, 2016ApJ...830...51S, 2018ApJ...858..100F}). Figure \ref{fig08} exhibits the distribution of four morphological subclasses and and four corresponding groups of galaxies in the {\it UVJ} diagram. The wedge-shaped region is dominated by quiescent galaxies, whereas galaxies in the remaining region are  recognized as star-forming galaxies.

The ETDs, LTDs, and IRR subclass mainly stay in the star-forming region.
However, the subclasses of ETDs and LTDs tend to invade into the quiescent region.
Most of the SPH galaxies are regarded as quiescence.  For the reason that quenching state is associated with bulge, our method can successfully strip the SPH galaxies out. As galaxy morphology transforms from IRR to SPH, it is found that galaxies tend to move into the wedged region. In general, it shows that there is a clear connection between {\it UVJ} colors and our morphological subclasses.  Our classification is reliable from this perspective.

\subsection{The distribution of morphological parameters}
In this subsection, we directly link our classification to morphological parameters
such as the S\'{e}rsic index, the effective radius, the Gini coefficient ($G$), and the second-order moment
of the 20\% brightest pixels ($M_{20}$).

\subsubsection{Parametric measurements}
Based on public {\it HST}/WFC3 F160W imaging,  \cite{2014ApJ...788...28V} measured S\'{e}rsic index $n$ and  effective radius $r_{\rm e}$ by modeling the galaxy as a single S\'{e}rsic profile \citep{1963BAAA....6...41S} using GALFIT \citep{2002AJ....124..266P}.

\begin{figure*}
\centering
\includegraphics[scale=1.0]{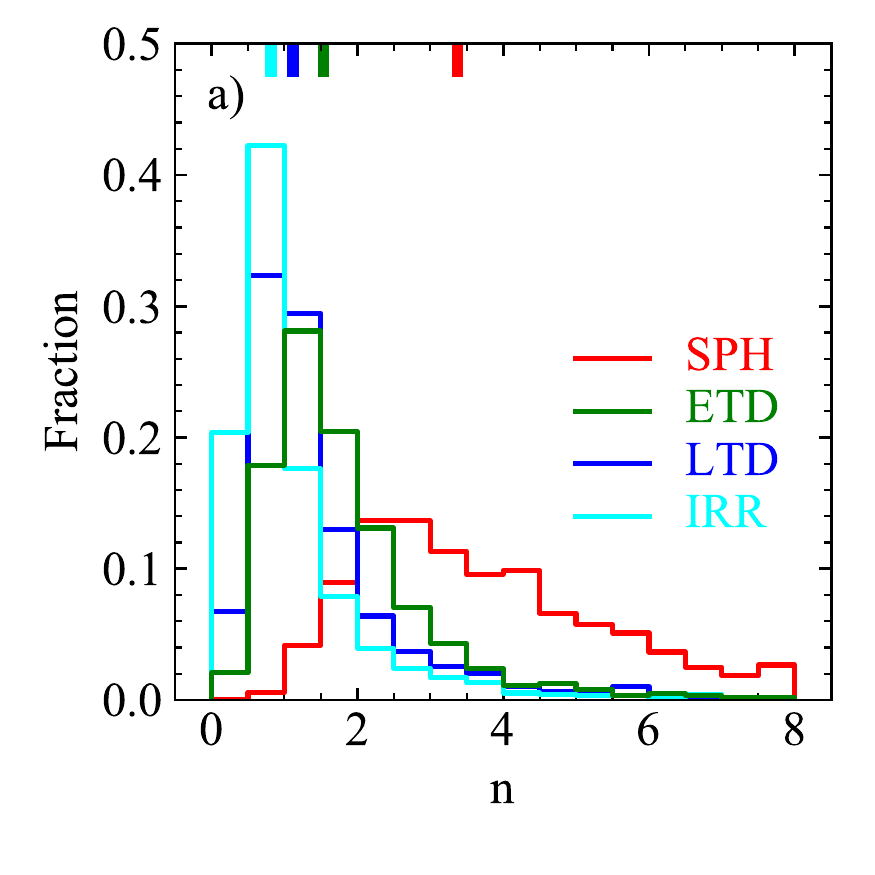}
\includegraphics[scale=1.0]{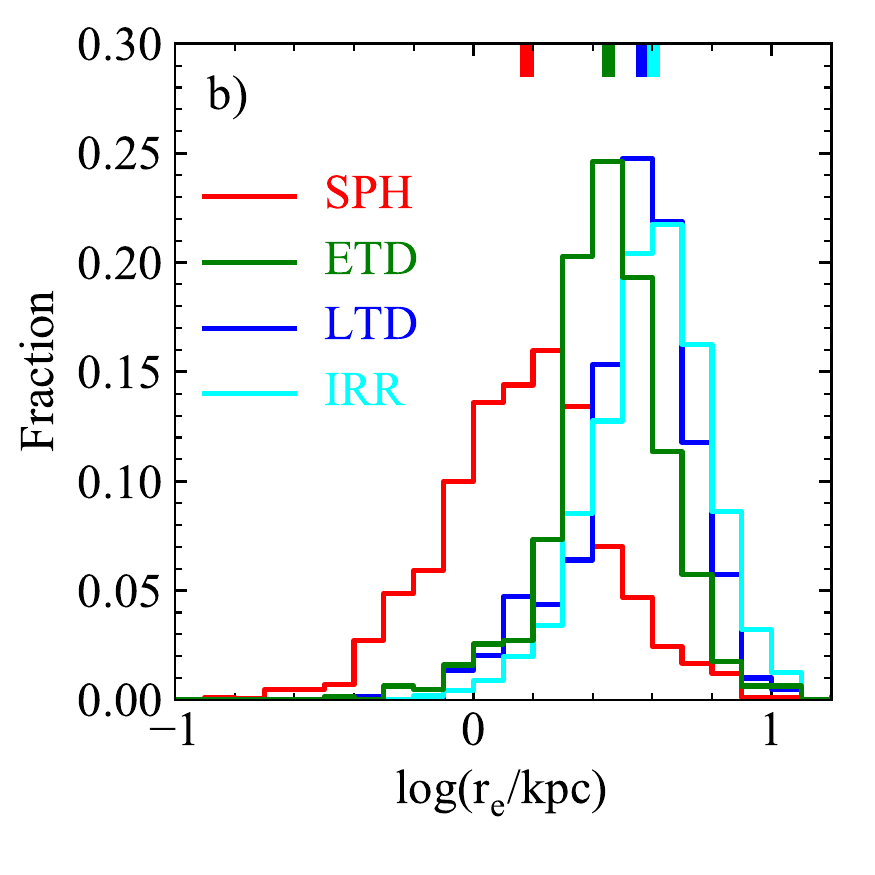}
\caption{Distributions of S\'{e}rsic index (left) and effective radius (right) for galaxies
that are merged into SPHs (red), ETDs (green), LTDs (blue), and IRRs (cyan). The median value of each subclass is denoted
by the upper bar with the corresponding color.
}
\label{fig09}
\end{figure*}

The histograms in the left panel of Figure \ref{fig09} show the distribution of S\'{e}rsic index for four subclasses of galaxies. The median S\'{e}rsic indexes of IRR, LTD, ETD, and SPH subclasses are 0.8, 1.1, 1.5, and 3.4, respectively. It is clear to see that there is an increasing trend of S\'{e}rsic index from IRRs to SPHs. The increasing sequence is highly in agreement with our common understanding of galaxy morphologies.
The histograms in the right panel of Figure \ref{fig09} show the distribution of effective radii for four subclasses of galaxies. The median effective radii of IRR, LTD, ETD, and SPH subclasses are 4.0, 3.7, 2.8, and 1.5 kpc, respectively. The SPH galaxies feature smaller size than other three subclasses. The typical size is a decreasing function following the sequence from IRRs to SPHs.

\subsubsection{Nonparametric measurements}
The nonparametric measurements are also introduced. We take the measurements of Gini and $M_{20}$
using the program Morpheus developed by \cite{2007ApJ...669..184A} on the $H$-band (F160W) images.
The Gini coefficient ($G$) is defined as \citep{2004AJ....128..163L}
\begin{equation}
    G = \frac{ \sum_{i}^{n}(2i - n- 1)\left| F_i \right|}{\left|\bar{F} \right|n(n-1)},
\end{equation}
where $F_i$ is the pixel flux value sorted in ascending order, $n$ is the total number of pixels
uniquely assigned to a galaxy during object detection, and $\bar{F}$ is a mean flux for all the pixels.
$M_{20}$  is defined as \citep{2004AJ....128..163L}
\begin{equation}
  M_{20} = \log\left(\frac{\sum_{i} M_i}{M_{tot}}\right) , {\rm with} \sum_i F_i < 0.2F_{\rm tot},
\end{equation}
where $M_i = F_i[(x_i - x_c)^2+(y_i - y_c)^2]$ and $M_{\rm tot} = \sum_{i=1}^{N}{M_i} $ for the fluxes
of the brightest 20\% of light in a galaxy.
In general, the Gini coefficient is a statistical coefficient to quantify the uniformity of light distribution,
while $M_{20}$ traces the substructures in a galaxy, such as bars, spiral arms, and multiple cores.

\begin{figure}
\centering
\includegraphics[scale=0.6]{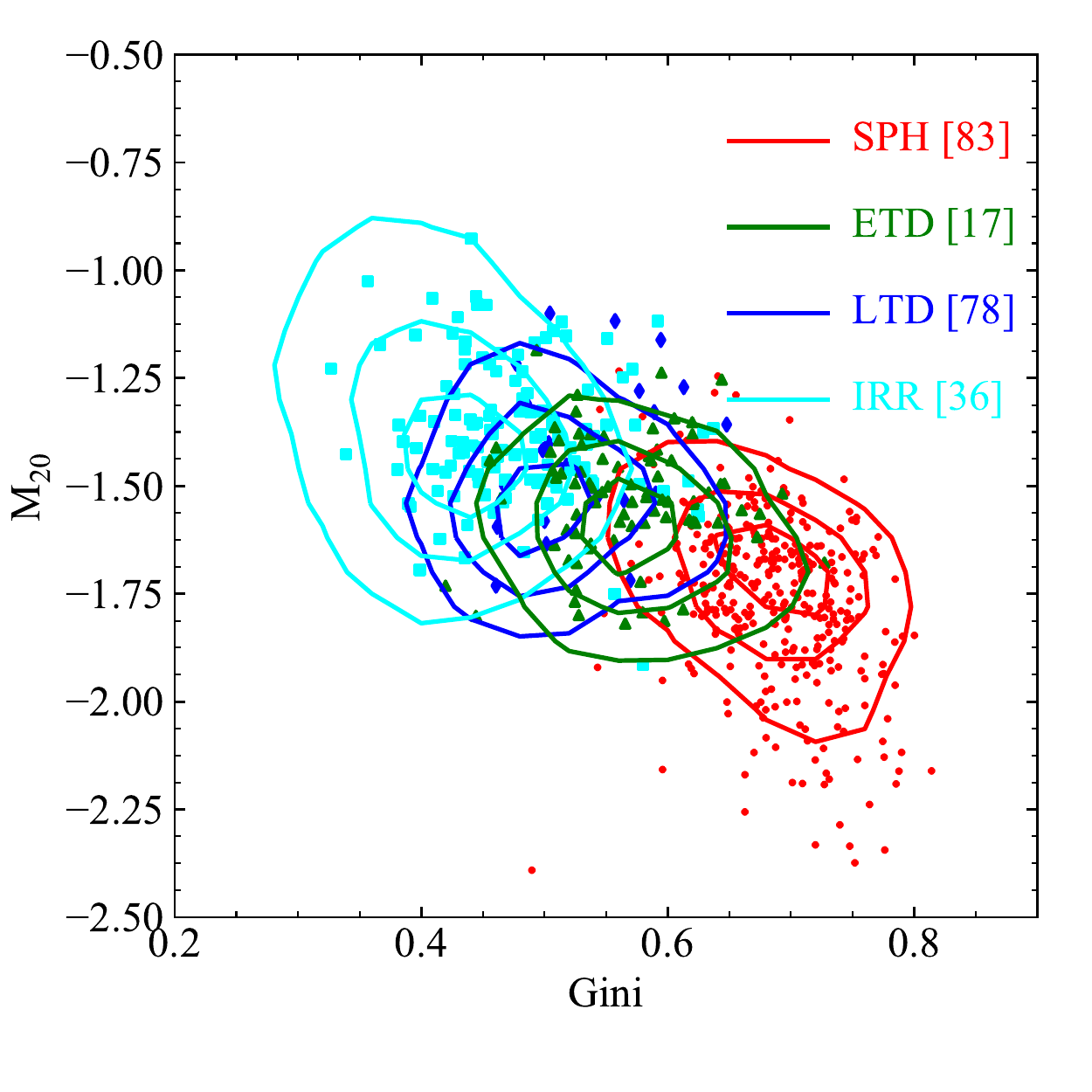}
\caption{Distributions of galaxies in the $G$--$M_{20}$ space for galaxies that are merged into SPHs (red), ETDs (green), LTD (blue), and IRR (cyan). The contour levels indicate the 20\%, 50\%, and 80\% of the corresponding subclass from inner to outskirts. The symbols are the same as in Figure \ref{fig08}.
}
\label{fig10}
\end{figure}

In Figure \ref{fig10}, we plot the distributions of four subclasses and four corresponding groups in the $G$--$M_{20}$ space. The sequence from IRRs to SPHs follows the orientation with increasing $G$ and decreasing $M_{20}$. The separation between
SPH and IRR galaxies in the $G$--$M_{20}$ plane is clearly represented here, where IRR galaxies predominantly
possess smaller $G$ and larger $M_{20}$ and SPH galaxies have larger $G$ and smaller $M_{20}$. The results
also prove that our classifications are closely aligned with nonparametric measurements.

\subsection{The influences of different $N$ and $dim_{\rm L6}$}
\label{subsec:diff_paras}

We also examine the influences of two adopted parameters during the learning on the final classifications: (1) $N$ in Equation (\ref{eq:normalization}) that describes the degree of discretization in the max--min normalization pretreatment, and (2) the number of features in the hidden-layer $dim_{\rm L6}$ (i.e., Layer 6 in Figure \ref{fig02} and Table \ref{tab_CAE}).

\begin{figure*}
\centering
\includegraphics[width=\textwidth]{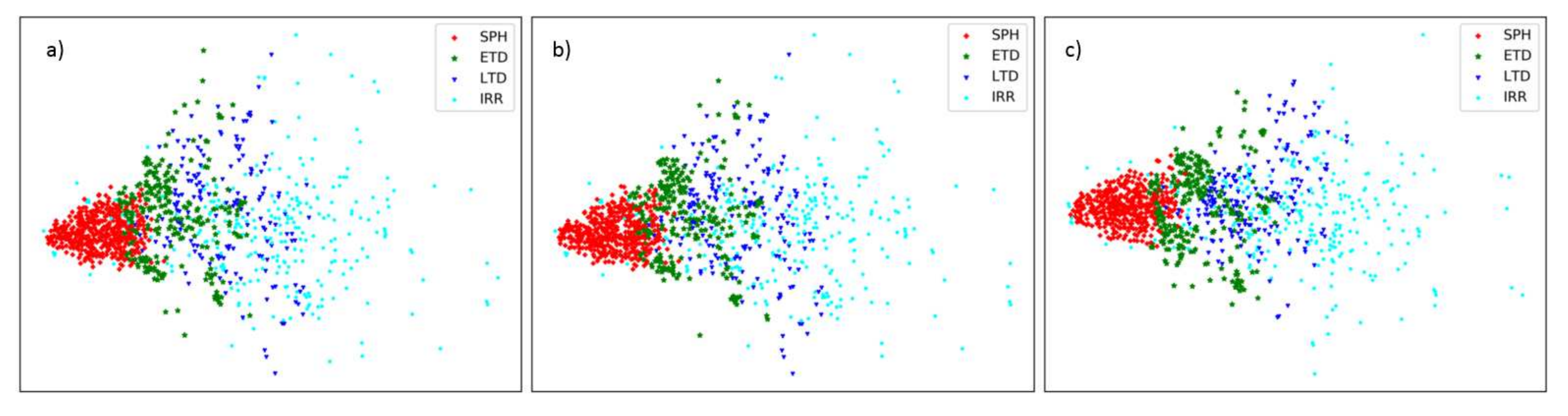}
\caption{Comparisons between different preprocessing settings. The t-SNE visualization graphs are based on the hidden layers that are encoded from data preprocessed by difference strategies. Panel (a) shows the results of our fiducial strategy with a degree of discretization of $N=500$ and a hidden-layer size of $dim_{\rm L6}=40$.
Panel (b) shows the results of the strategy with $N=700$ and $dim_{\rm L6}=40$.
Panel (c) shows the results of the strategy with $N=700$ and $dim_{\rm L6}=100$.
It shows that our strategy is reasonable, since no significant improvements are given by other strategies.}
\label{fig11}
\end{figure*}

In this work, we adopt $N=500$ and $dim_{\rm L6}=40$ as fiducial values, the resulting t-SNE visualization graph for the four main subclasses is given in Figure \ref{fig11}, together with results based on different adopted values of these parameters (i.e., $N=700$ and $dim_{\rm L6}=100$). The comparison between results derived from different combinations of the these two parameters indicate that the choice of a larger latent size of 100 or a larger degree of discretization of 700 gives no significant improvements. On the other hand, in Section \ref{subsec:dim_cae} we have demonstrated that a hidden layer with feature number of 40 is enough to recover most of the morphological characteristics for our sample. Thus, our fiducial values of these two parameters are reasonable.

\subsection{Comparisons with visual classification and the supervised method}
\label{subsec:comp_5tag}

In this section, we compare our results with other works in which galaxies are broadly split into the 5-tag case: SPH, ELD, LTD, IRR, and UNC.
In the program of CANDELS visual classifications \citep{2015ApJS..221...11K},
$\sim8000$ galaxies in the GOODS-S field are classified by human visualization, where one image would be labeled by independent classifiers. Using these labeled galaxies as the training dataset, \cite{2015ApJS..221....8H} proposed a supervised deep-learning method and then applied to other four fields. To define the morphological class, five parameters (i.e., $f_{\rm spheroid}, f_{\rm disk}, f_{\rm irr}, f_{\rm PS}, f_{\rm Unc}$) for each galaxy are produced and retrieved through analysis of its $H$-band image.  The definition of galaxy classifications is shown as follows (see  \citealt{2015ApJS..221....8H}):
\begin{enumerate}
\item Spheroids (SPH): $f_{\rm spheroid} > 2/3$, $f_{\rm disk} < 2/3$, and $f_{\rm irr} < 0.1$ ;
\item Early-type Disks (ETD): $f_{\rm spheroid} > 2/3$, $f_{\rm disk} > 2/3$, and $f_{\rm irr} < 0.1$ ;
\item Late-type  Disks (LTD): $f_{\rm spheroid} < 2/3$, $f_{\rm disk} > 2/3$, and $f_{\rm irr} < 0.1$;
\item Irregulars (IRR): $f_{\rm spheroid} < 2/3$ and $f_{\rm irr} > 0.1$.
\item Unclassifiable (UNC): the remaining
\end{enumerate}
\normalsize

\begin{figure*}
\centering
\includegraphics[scale=0.22]{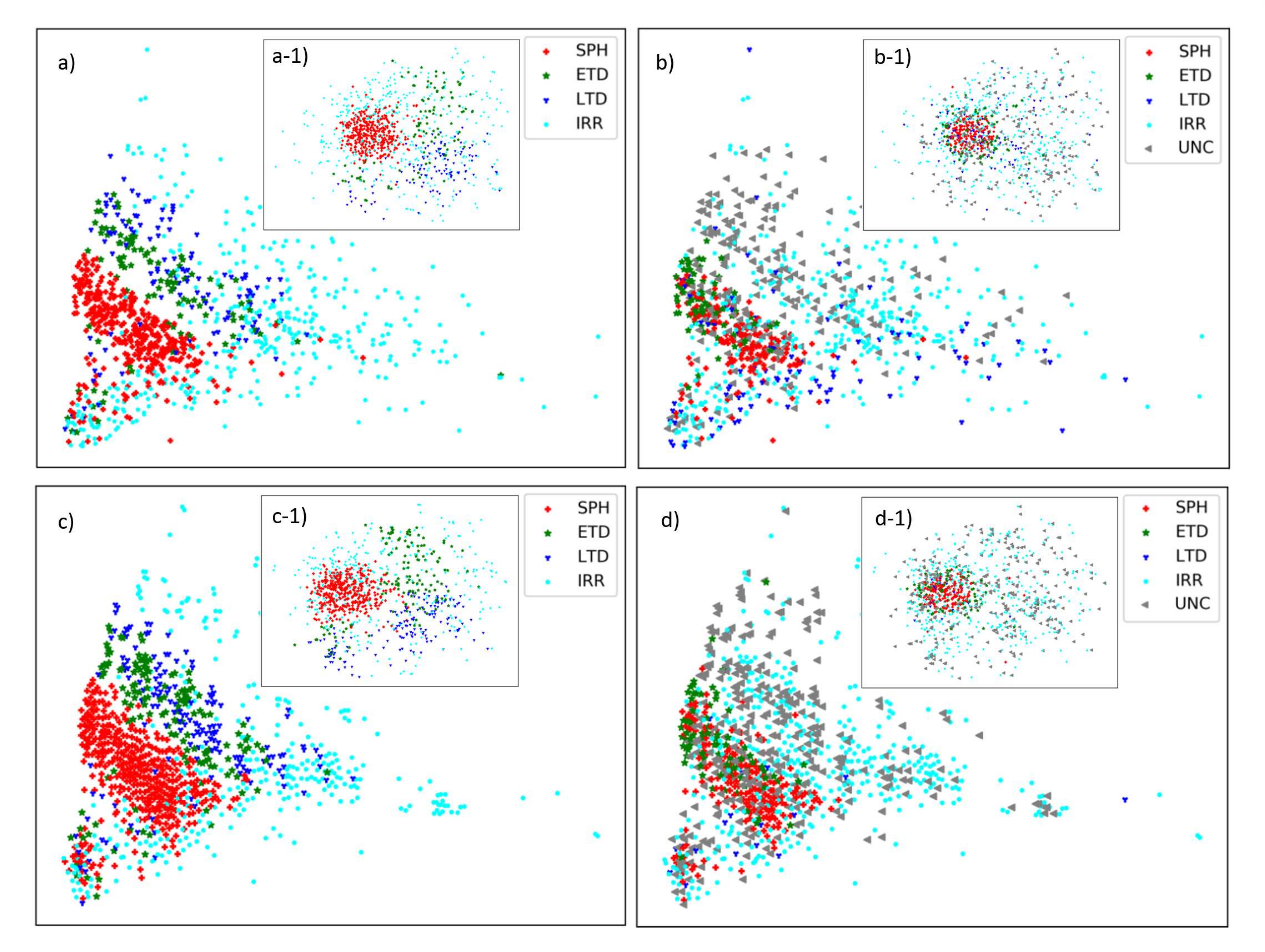}
\caption{
The t-SNE visualization graphs based on raw images of the randomly selected subsamples. The embedded subdiagrams are the t-SNE visualization graphs based on the encoded hidden features of the CAE. The upper panels show the randomly selected galaxies from all the five CANDELS fields. Panels a and b are color coded by our results and those of the supervised method \citep{2015ApJS..221....8H}, respectively. The bottom panels show the randomly selected subsamples only from the GOODS-S field. Panels c and d are color coded by our results and those of CANDELS visual classifications \citep{2015ApJS..221...11K}, respectively. It shows that our results (i.e., Panels a and c), have more discrete distributions.}

\label{comp_zoom_2015}
\end{figure*}

We cross match our sample with morphology catalogs released by these two works and perform direct comparisons via the t-SNE map in Figure \ref{comp_zoom_2015}. For the supervised deep-learning results of \cite{2015ApJS..221....8H}, 1000 matched galaxies from the CANDELS five fields are randomly selected and plotted in the t-SNE map based on the raw images, color coded by our classifications (Panel a) and those of \cite{2015ApJS..221....8H} (Panel b).
For the visual classifications of \cite{2015ApJS..221...11K}, since only catalog for GOODS-S field was released, we randomly select 1000 matched galaxies from the same field and exhibit the raw-image-based t-SNE map color coded by our classifications in Panel c, while the same map for the results from \cite{2015ApJS..221...11K} are given in Panel d. In each panel,
the t-SNE graph based on the final hidden-layer is also provided in the inset as a supplement.

Since our results are not entirely consistent with those of previous studies, it is possible that some galaxies are classified as the UNC category in other works, as shown in Panels b and d of Figure \ref{comp_zoom_2015}. The computation of the t-SNE maps based on either the raw images or the encoded hidden-layer naturally contains (nearly) all the morphological information of the galaxies used, thus one might expect that for a better classification, galaxies belonging to the same morphological type should be apparently more clustered, while the distributions of galaxies from different types should be more distinguishable and less mixed. Based on this simple and plausible deduction, the following conclusions can be reached by the t-SNE maps shown in Figure \ref{comp_zoom_2015}: (1) the similarity between Panels b and d suggests that the supervised deep-learning method developed by \cite{2015ApJS..221....8H} performs well using the training set from \cite{2015ApJS..221...11K}, and (2) however, the classification results from
our method generally performs better in identifying galaxies belonging to different morphological types than both of the above works, mainly due to the clearer clustering of galaxies from the same subclass.

\subsection{Comparison with other UML method referenced by the supervised results}

\cite{2018MNRAS.473.1108H} applied a UML technique that combines the growing neural gas
algorithm \citep{Fritzke95} and the hierarchical clustering technique to automatically
segment and label galaxies. Because galaxies in \cite{2018MNRAS.473.1108H} are clustered
into 100--200 groups, we choose the 100-group case for an equivalent comparison with our 100-tag results
(24,900 galaxies that are well-clustered).
Given that the direct mapping between these results may not exist, we take the 5-tag
classifications of \cite{2015ApJS..221....8H} as reference to see how consistent between the
results from these two UML methods and the one from the supervised deep learning.
However, we are not intending to claim a better classification scheme of \cite{2015ApJS..221....8H} than the other two, since in Section \ref{subsec:comp_5tag} we have demonstrated that our morphological results perform better than those of \cite{2015ApJS..221....8H}
in term of clustering galaxies belonging to the same morphology in the t-SNE maps.

\begin{figure*}
\centering
\includegraphics[scale=0.8]{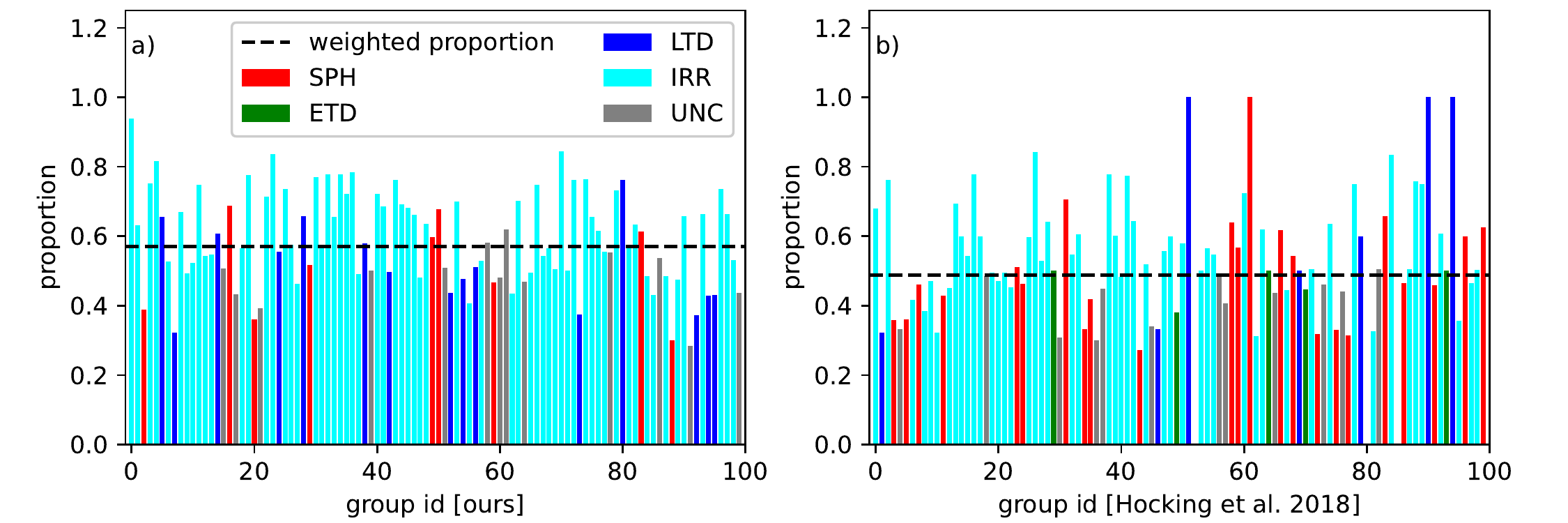}

\caption{Proportion of the dominated class based on the \cite{2015ApJS..221....8H} classifications for 100 groups of our sample from our results (left) and those of \cite{2018MNRAS.473.1108H} (right). Each group is color coded by its dominated class, i.e., SPH (red), ETD (green), LTD (blue), IRR (cyan), or UNC (gray). The black dashed lines in each panel denote the weighted proportion of all the 100 groups.}
\label{100-our-2015}
\end{figure*}

We use our well-clustering sample (100 groups containing SPH, ETD, LTD, IRR, and UNC) to draw the following comparison. These galaxies can be matched by the object ID from the 3D-HST catalogs. We cross match our sample with those of \cite{2015ApJS..221....8H}, and for each group of our unsupervised method, we can find a dominated class of the \cite{2015ApJS..221....8H} classifications (i.e., SPH, ELD, LTD, IRR or UNC). The proportions of the dominated class for each of the 100 groups are computed and illustrated in the left panel of Figure \ref{100-our-2015}, color coded by the dominated classes of the groups. For our well-clustering sample, similar plot is computed for the 100-tag results of \cite{2018MNRAS.473.1108H} and is shown in the right panel of Figure \ref{100-our-2015}.  The proportion weighted by the number in each group, which can be used to quantify the consistency between the classifications of \cite{2015ApJS..221....8H} and another ones, is $\sim$57\% for our results, higher than $\sim$49\% for the
results of \cite{2018MNRAS.473.1108H}, suggesting that our morphological results are more consistent with those of the supervised method. We thus conclude that the voting strategy described in Section \ref{subsec:voting} improves the quality of clustering galaxies at the cost of the discard sources.

\section{Conclusions}
\label{sec:sum}

In this paper, we present a UML method that can automatically classify galaxies with similar morphology
in deep filed surveys. The method consists of two steps: (1) the CAE is used to compress the dimensions
from the raw data and extracts features, and (2) the bagging-based multiclustering method sorts out the assuring galaxies with analogous characteristics into one group. After discarding the galaxies with inconsistent results of voting,  the remaining galaxies are well-clustered into 100 groups.

To further investigate our galaxy morphologies, we merge the groups into five main categories (SPH, ELD, LTD, IRR, and UNC) by visual verification. After discarding the low-S/N UNC category, we utilize massive galaxies ($M_*>10^{10}M_\odot$) to explore the robustness of the morphological classifications by the connection with other physical properties. We show that this classification scheme is in a good agreement with other galaxy properties, including {\it UVJ} diagnoses and other morphological parameters.
The comparison suggests that our method gives a robust result of morphological classification. Overall,
the unsupervised method provides an independent feature extraction and galaxy classification only utilizing
the monochromatic H-band images. It is able to obtain the reasonable clustering of galaxies with similar features first,
and further increasing the efficiency of visual classification.

Since a strict voting model is applied, the clustering subsamples with high
quality can be used as a training sample in other downstream tasks such as the supervised machine learning.
In the future, we intend to develop the techniques for multicolor images and apply them
to the data processing from the Euclid and CSST. The techniques provide the results of galaxy classifications
with few human intervention. We expect that the future work would help us to have a better understanding
of the morphologies of galaxies as well as their formation and evolution.

\acknowledgments
This work is based on observations taken by the 3D-HST
Treasury Program (GO 12177 and 12328) with the NASA/
ESA HST, which is operated by the Association of Universities for Research in Astronomy, Inc., under NASA contract NAS526555.
This work is supported by the National Natural Science Foundation of China (NSFC; Nos. 11673004, 62106033)
and the National Basic Research Program of China (973 Program; 2015CB857004). C.C.Z. acknowledges the support from  Yunnan Youth
Basic Research Projects (202001AU070020).
Y.Z.G. acknowledges the support from China Postdoctoral Science Foundation (2020M681281) and Shanghai Post-doctoral Excellence Program (2020218). G.W.F. acknowledges the support from  Yunnan Applied Basic Research Projects (2019FB007). Z.S.L. acknowledges the support from China Postdoctoral Science Foundation (2021M700137). We acknowledge the science research grants from the China Manned Space Project with NO. CMS-CSST-2021-A07.


\bibliographystyle{aasjournal}
\bibliography{ref}

\end{document}